\documentstyle[epsf]{article}

\newcommand{\gsim}{\;\rlap{\lower 3.5 pt \hbox{$\mathchar \sim$}} \raise 1pt
 \hbox {$>$}\;}
\newcommand{\lsim}{\;\rlap{\lower 3.5 pt \hbox{$\mathchar \sim$}} \raise 1pt
 \hbox {$<$}\;}

\begin{document}

\title{
\vskip-3cm{\baselineskip14pt
\centerline{\normalsize\hfill MPI/PhT/97--79}
\centerline{\normalsize\hfill hep-ph/9711465}
\centerline{\normalsize\hfill November 1997}
}
\vskip1.5cm
Higher Order Corrections to the Hadronic Higgs Decay\thanks{
To appear in {\it Proceedings
of the International Europhysics Conference on High Energy Physics},
Jerusalem, Israel, 19-26 August 1997.
}
}
\author{Matthias\, Steinhauser
}
\date{}
\maketitle

\vspace{-1em}
\begin{center}
Max-Planck-Institut f\"ur Physik,
    Werner-Heisenberg-Institut,\\ D-80805 Munich, Germany
\end{center}
\vspace{.3cm}

\begin{abstract}
\noindent
A Higgs boson in the intermediate mass range is considered. An
effective Lagrangian approach is used in order to
evaluate the top-induced QCD corrections of order $\alpha_s^3$ 
to the decay into light quarks and ${\cal O}(\alpha_s^4)$ corrections
to the gluonic decay mode.
The connection to the decoupling relations for $\alpha_s$ 
and the light quark masses is discussed.
\end{abstract}

\vspace{.3cm}

The Higgs boson is the only still missing particle in the 
standard model. Currently only lower bounds of roughly
$70$~GeV have been obtained from the
failure of finding the Higgs boson at the Large Electron-Positron
Collider (LEP) at CERN.
A global fit to the precision date also prefers a Higgs mass
in the so-called intermediate mass range with $M_H\lsim2M_W$.
In this contribution we will concentrate on such a Higgs boson.
The dominant decay mode is then the one into bottom quarks.
In the ``massless'' theory corrections are known up to 
${\cal O}(\alpha_s^3)$~\cite{Che97}. In this contribution
we will discuss the top-induced correction terms of the same 
order~\cite{CheSte97}.

A second very interesting decay mode of a Higgs boson in the
intermediate mass range is the decay into gluons. The lowest
order process is mediated by a quark loop.
It is interesting to note that in the limit where the internal
quark mass is much larger than $M_H$ the amplitude gets independent
of the quark mass. Therefore this process counts the number of 
heavy quarks. In contrast to the electroweak $\rho$ parameter \cite{vel}, 
the $ggH$ coupling is also sensitive to quark isodoublets if they 
are mass-degenerate.
The first order QCD corrections to the triangle diagram where a top quark
is running around are known since long~\cite{ina,daw,djo}.
It turned out that the order $\alpha_s$ corrections amount to
approximately $70$~\%. 
These large corrections were a strong motivation to evaluate the 
next term in the expansion in $\alpha_s$~\cite{CheKniSte97hgg}.
Note that the $ggH$ coupling also appears as a building
block in the process 
$gg\to H$ which is the dominant production mechanism at LHC.

As the top quark is much heavier than all other mass scales involved
in the processes under consideration it makes sense to construct in a
first step an effective Lagrangian.
We start with 
the bare Yukawa Lagrangian,
\begin{equation}
{\cal L}_Y = -\frac{H^0}{v^0}
\left(
 \sum_q m_q^0 \bar{q}^0 q^0 + m_t^0 \bar{t}^0 t^0
\right)
{},
\end{equation}
where $v$ is the Higgs vacuum-expectation value and
the superscript 0 labels bare quantities.
Assuming that the Higgs boson  mass, $M_H$, is less than the
top quark mass, $m_t$, ${\cal L}_Y$  can be replaced by an 
effective Lagrangian  produced by integrating out  the top
quark field.  According to Refs.~\cite{ina,CheKniSte97}
the resulting Lagrangian reads
\begin{eqnarray}
{\cal L}_Y^{\rm eff} &=& -\frac{H^0}{v^0}\left[
C_1 \left[O_1^\prime\right]
+\sum_q
C_{2} \left[O_{2q}^\prime\right]
\right],
\label{eqlageff}
\end{eqnarray}
where $[O_1^\prime]$ and $[O_{2q}^\prime]$ are the 
renormalized counterparts of 
the bare operators
\begin{eqnarray}
O_1^\prime\,\,=\,\,\left(G_{a\mu\nu}^{0\prime}\right)^2,
&&
O_{2q}^\prime\,\,=\,\,m_q^{0\prime}\bar q^{0\prime}q^{0\prime},
\nonumber
\end{eqnarray}
with $G_{a\mu\nu}^{0\prime}$ being the (bare) field strength tensor of
the gluon.  The primes mark the quantities defined in the effective
$n_f = 5 $ QCD including only light quarks. 
All the dependence on the top quark gets localized in
the coefficient functions $C_1$ and $C_{2}$.

The task to compute the decay rates essentially splits into
two parts: $(i)$ The evaluation of the coefficient functions where the scale
is given by the top quark mass, and 
$(ii)$ the computation of the imaginary part of the correlators formed
by the operators $O_1^\prime$ and $O_{2q}^\prime$. There the particles
inside the loops are massless and for the external momentum one has 
$q^2=M_H^2$.

\begin{figure}[t]
 \begin{center}
 \begin{tabular}{c}
   \leavevmode
   \epsfxsize=3.5cm
   \epsffile[189 314 481 531]{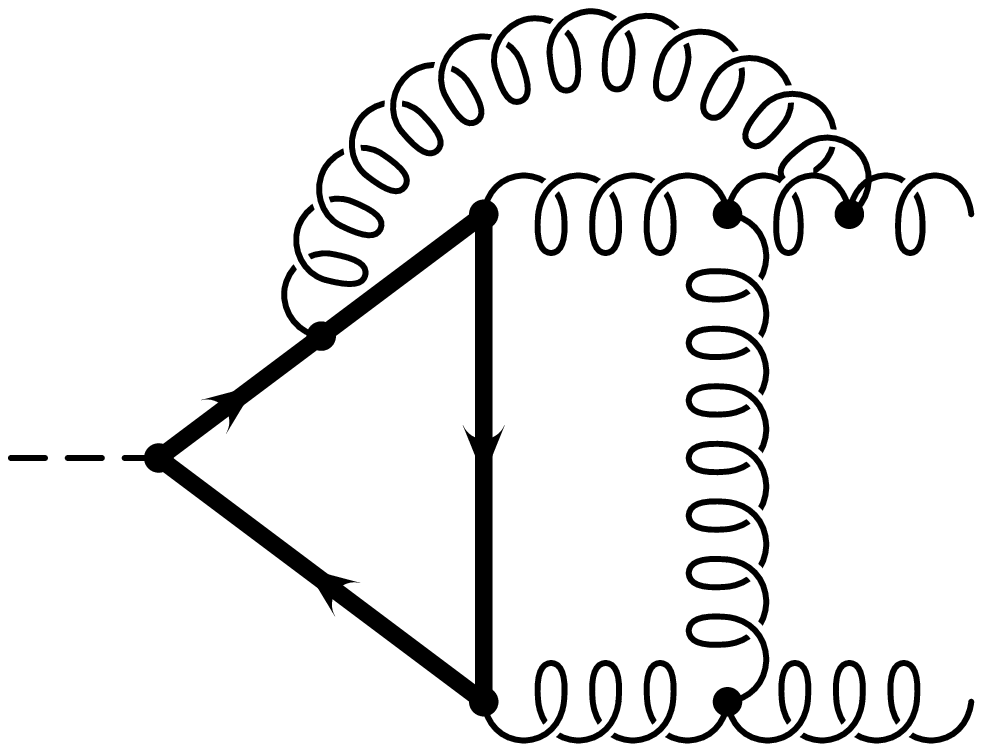}
   \hphantom{xxxxx}
   \epsfxsize=3.5cm
   \epsffile[189 350 481 531]{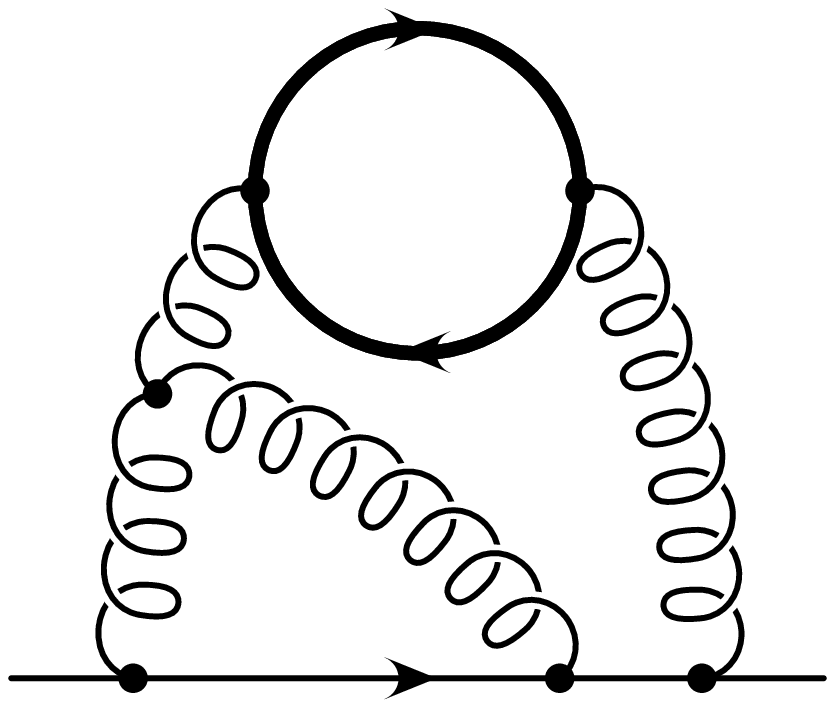}
   \\
   \epsfxsize=3.5cm
   \epsffile[189 314 481 531]{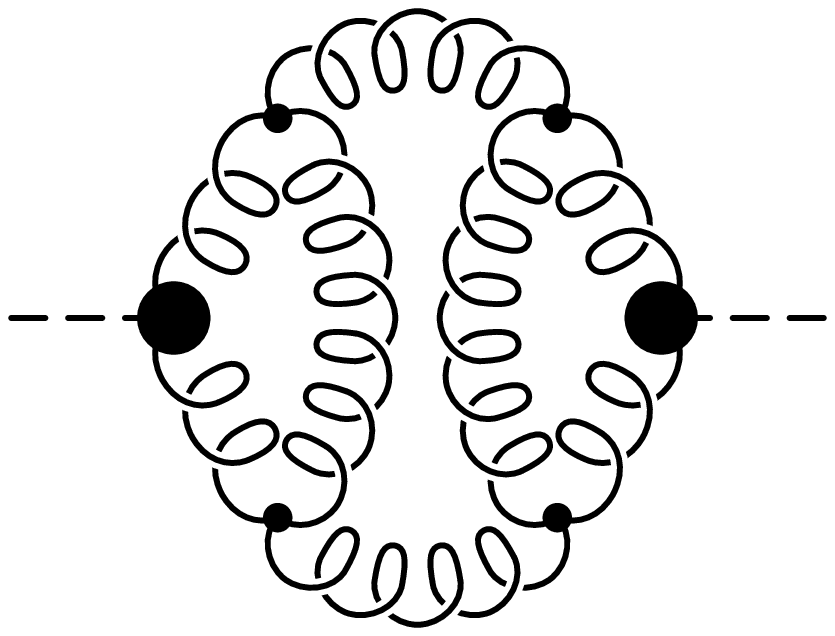}
   \epsfxsize=3.5cm
   \epsffile[189 314 481 531]{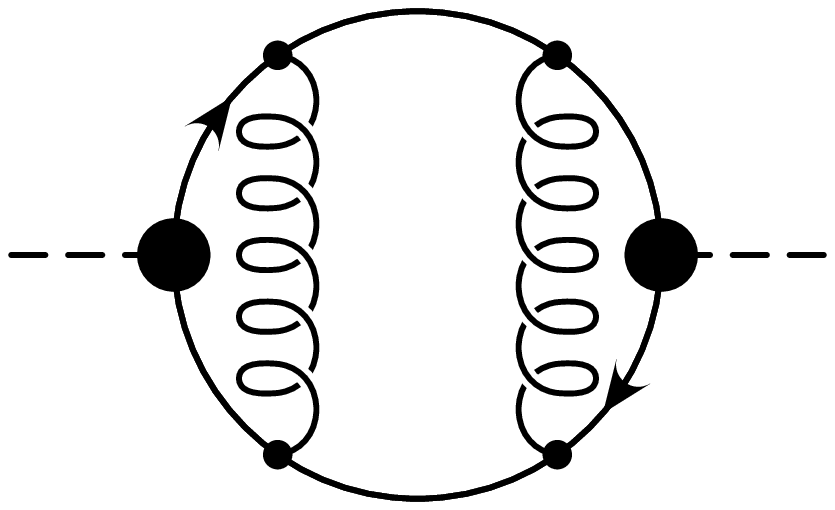}
   \epsfxsize=3.5cm
   \epsffile[189 314 481 531]{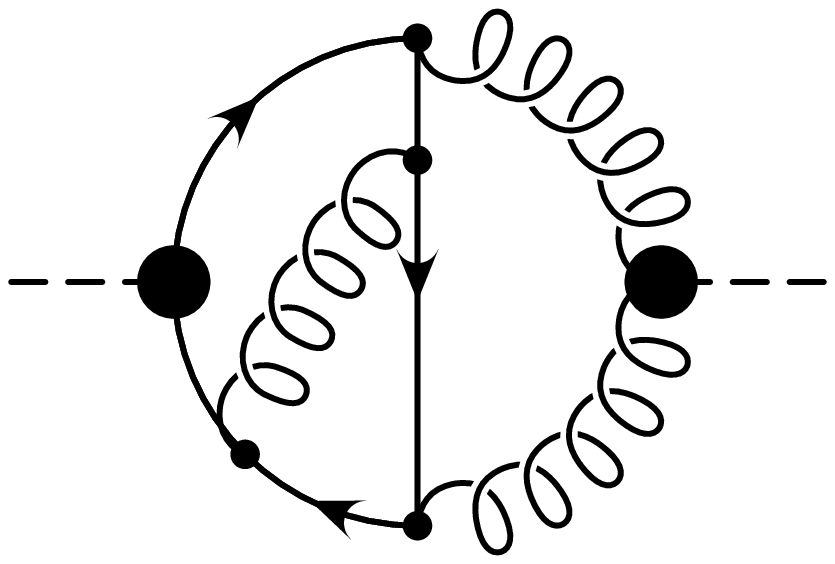}
 \end{tabular}
\caption{\label{figdia}
         Typical Feynman diagrams contributing to $C_1$, $C_2$
         and the three kind of correlators.}
 \end{center}
\end{figure}

The computation of the coefficient function $C_1$ has been
performed in two different way. 
The first method considers the vertex diagrams where the Higgs
boson couples the top quark and where two gluons are in the final state
(see Fig.~\ref{figdia}).
In order to calculate $C_1$ these graphs have to be expanded in their
two external momenta. 

The second method is based on a low energy theorem (LET). We were
able to derive the following simple formulae~\cite{CheKniSte97dec}:
\begin{eqnarray}
C_1 = - \frac{1}{2} \frac{\partial\ln\zeta_g^2}{\partial\ln m_t^2},
&\hphantom{xxxx}&
C_2 = 1+2\frac{\partial\ln\zeta_m}{\partial\ln m_t^2},
\label{eqc12let}
\end{eqnarray}
where 
$\zeta_g$ and $\zeta_m$ are the decoupling constants relating
the strong coupling constant, $\alpha_s$, and the light quark masses,
$m_q$, in the full and effective theory:
\begin{eqnarray}
\alpha_s^\prime = \zeta_g\alpha_s,
&\hphantom{xxxx}&
m_q^\prime = \zeta_m m_q.
\end{eqnarray}
$\zeta_g$ may be computed by considering the diagrams containing
at least one top quark line of the 
gluon and ghost propagator and the gluon-ghost vertex with
external momentum equal to zero.
$\zeta_m$ is given by top-induced diagrams contributing to
the fermion propagator (see Fig.~\ref{figdia}). 
In~\cite{CheKniSte97alphas,CheKniSte97dec} 
$\zeta_g$ and $\zeta_m$ were evaluated up to the three-loop order.
Eqs.~(\ref{eqc12let}) reduce the computation of $C_1$ and $C_{2}$ 
to the evaluation of essentially two-point functions with 
vanishing external momentum.
With this method there are less diagrams to be considered which are
in addition much simpler to evaluate.
Another nice feature of Eqs.~(\ref{eqc12let}) is that because of the
logarithmic derivative even the ${\cal O}(\alpha_s^4)$ contributions to
$C_1$ and $C_{2}$ can be evaluated if $\zeta_g$ and $\zeta_m$
are known up to ${\cal O}(\alpha_s^3)$.

There are actually three types of correlators which have to be evaluated.
In the bottom line of Fig.~\ref{figdia} some sample diagrams are pictured.
The computation was performed with the program package
MINCER~\cite{MINCER}.

The combination of the coefficient function and the imaginary part
of the correlator
$\langle\left[O_1^\prime\right]\left[O_1^\prime\right]\rangle$
immediately leads to the decay rate $\Gamma(H\to gg)$ ($\mu=M_H$):
\begin{eqnarray}
\label{fin}
\lefteqn{\frac{\Gamma(H\to gg)}{\Gamma^{\rm Born}(H\to gg)}
=}
\nonumber\\&&
1+
\frac{\alpha_s^{(5)}(M_H)}{\pi}\left(\frac{95}{4}-\frac{7}{6}n_l\right)
+\left(\frac{\alpha_s^{(5)}(M_H)}{\pi}\right)^2
\left[\frac{149533}{288}-\frac{363}{8}\zeta(2)
\right.\nonumber\\
&&\left.\mbox{}
-\frac{495}{8}\zeta(3)
-\frac{19}{8}\ln\frac{M_t^2}{M_H^2}
+n_l\left(-\frac{4157}{72}+\frac{11}{2}\zeta(2)+\frac{5}{4}\zeta(3)
-\frac{2}{3}\ln\frac{M_t^2}{M_H^2}\right)
\right.\nonumber\\
&&\left.\mbox{}
+n_l^2\left(\frac{127}{108}-\frac{1}{6}\zeta(2)\right)\right]
\nonumber\\
&\approx&1+17.917\,\frac{\alpha_s^{(5)}(M_H)}{\pi}
+\left(\frac{\alpha_s^{(5)}(M_H)}{\pi}\right)^2
\left(156.808-5.708\,\ln\frac{M_t^2}{M_H^2}\right),
\end{eqnarray}
with $\Gamma^{\rm Born}(H\to gg)
      =G_FM_H^3/36\pi\sqrt2 \times (\alpha_s^{(5)}(M_H)/\pi)^2$.
Choosing $M_t=175$~GeV and $M_H=100$~GeV one arrives at:
\begin{eqnarray}
\label{num}
\frac{\Gamma(H\to gg)}{\Gamma^{\rm Born}(H\to gg)}
&\approx&1+17.917\,\frac{\alpha_s^{(5)}(M_H)}{\pi}
+150.419\,\left(\frac{\alpha_s^{(5)}(M_H)}{\pi}\right)^2
\nonumber\\
&\approx&1+0.66+0.21.
\end{eqnarray}
We observe that the new ${\cal O}(\alpha_s^2)$ term further increases the
well-known ${\cal O}(\alpha_s)$ enhancement by about one third.
If we assume that this trend continues to ${\cal O}(\alpha_s^3)$ and beyond,
then Eq.~(\ref{fin}) may already be regarded as a useful approximation to the
full result.
Inclusion of the new ${\cal O}(\alpha_s^2)$ correction leads to an increase of 
the Higgs-boson hadronic width by an amount of order 1\%.

The decay rate $\Gamma(H\to b\bar{b})$ can be cast into the form
\begin{eqnarray}
\lefteqn{
\Gamma(H\to b\bar{b}) = 
A_{b\bar{b}}\Bigg\{
1
+ 5.667 \,  a_H^{(5)}
+ 29.147 \, \left(a_H^{(5)}\right)^2
+ 41.758 \left(a_H^{(5)}\right)^3
}
\label{eqbb}
\\&&\mbox{}
+\left(a_H^{(5)}\right)^2
\left[
3.111
-0.667\,L_t
\right] 
+\left(a_H^{(5)}\right)^3
\left[
50.474
-8.167\,L_t
-1.278\,L_t^2
\right]
\Bigg\},
\nonumber
\end{eqnarray}
with $A_{b\bar b}=3G_FM_Hm_b^2/4\pi\sqrt{2}$,
$L_t=\ln M_H^2/M_t^2$ and $a_H^{(5)}=\alpha_s^{(5)}(M_H)/\pi$.
In Eq.~(\ref{eqbb}) electromagnetic and electroweak
corrections have been neglected. Also mass correction terms
and second order QCD corrections which are suppressed by the 
top quark mass are not displayed.
One observes from Eq.~(\ref{eqbb}) that the top-induced corrections
at ${\cal O}(\alpha_s^3)$ are of the same order of magnitude
than the ``massless'' corrections.

I would like to thank K.G. Chetyrkin and B.A. Kniehl
for the fruitful collaboration on this subject.

%

\end{document}